\documentclass[final,5p,times,twocolumn]{elsarticle}

\begin{document}

	\begin{frontmatter}



\title{Rattling phonon modes in quadruple perovskites}



\author{Z.\,V.\, Pchelkina$^{a,b}$,
E.\,V.\, Komleva$^{a,b}$, 
V.\,Yu.\, Irkhin$^{a}$, 
Y.\, Long$^{c,d}$, 
S.\,V.\, Streltsov$^{a,b}$\/\thanks{e-mail: streltsov@imp.uran.ru}}



\address{$^a$Institute of Metal Physics, S. Kovalevskaya Street 18, 620108 Ekaterinburg, Russia\\~\\
$^b$Department of Theoretical Physics and Applied Mathematics, Ural Federal University,
Mira St. 19, 620002 Ekaterinburg, Russia\\~\\
$^c$Institute of Physics, Chinese Academy of Sciences, 100190 Beijing, China\\~\\
$^d$Songshan Lake Materials Laboratory, Dongguan, Guangdong 523808, China}

\begin{abstract}
{
Rattling phonon modes are known to be origin of various anomalous physical properties such as superconductivity, suppression of thermal conductivity, enhancement of specific heat etc. By means of DFT+$U$ calculations we directly show presence of the 
rattling mode in the quadruple perovskites CuCu$_3$V$_4$O$_{12}$ and  CuCu$_3$Fe$_2$Re$_2$O$_{12}$ and argue that this can develop in others as well. It is demonstrated that Cu ions at $A$ sites vibrate in the center of the icosahedral oxygen O$_{12}$ cages and the corresponding potential has a complicated form with many local minima. 
}

\end{abstract}
	\end{frontmatter}




\section{Introduction}

 Effects of strong anharmonicity of the lattice potential in perovskites, including layered cuprates, three-dimensional perovskites and related systems are widely discussed \cite{IKT,Menushenkov,Akizuki2015}. Moreover,  oxygen atoms at the corners of the CuO$_6$ octahedra in cupric oxide (CuO) are supposed to be in a double-well potential. This fact was confirmed  for a number of high-$T_c$ superconductors and related parent systems, including YBa$_2$Cu$_3$O$_{7-\delta}$, La$_{2-x}$Sr$_x$CuO$_4$, and Nd$_{2-x}$Ce$_x$CuO$_{4-\delta}$ by the extended X-ray absorption fine structure (EXAFS)  experiments,  M\"{o}ssbauer investigations, and first-principle total energy calculations (see \cite{IKT,Menushenkov,IKT1,Menushenkov1} and references therein).  Similar situation with anomalous oxygen vibrations was observed in superconducting Ba$_{1-x}$K$_x$BiO$_3$ 
\cite{Men2}. In Ref.\cite{Bishop} appearance of the double-well potential in superconducting La$_{2}$CuO$_4$ was explained in terms of the Jahn-Teller polaron model. The double-well potential in double perovskites was discussed in Ref. \cite{Roy}, calculations of zone-center soft modes being carried out in order to characterize the polar and octahedral-rotation instabilities. Phonon modes in such potentials can be rather unusual.

\begin{figure}[h]
\includegraphics[width=0.8\columnwidth]{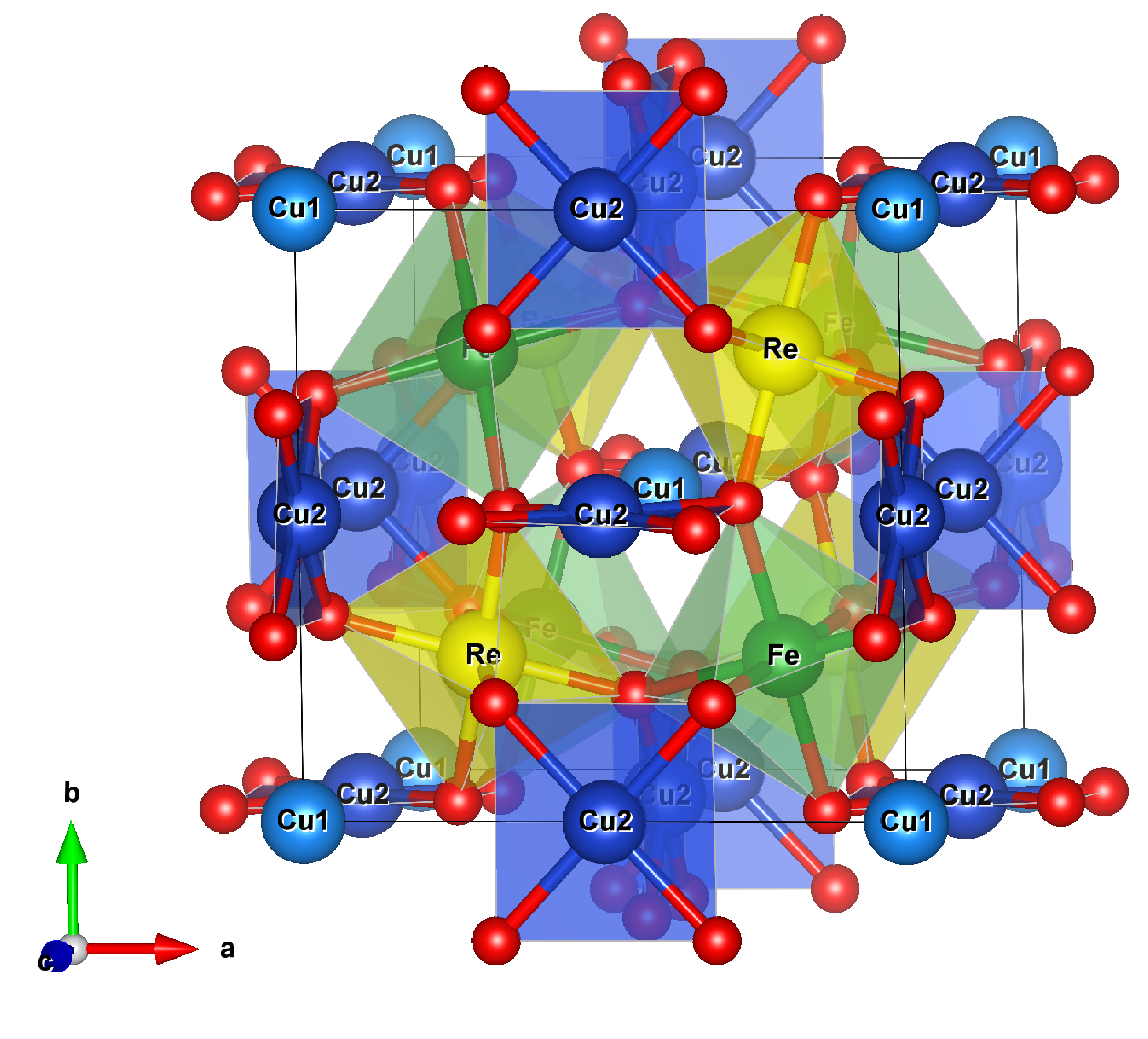}
\caption{\label{cucufere_str} Crystal structure of quadruple perovskites  A'$_3$B$_4$O$_{12}$ on example of CuCu$_3$Fe$_2$Re$_4$O$_{12}$. There are two types of Cu ions: Cu1 (light blue) is in a large cage, $A$ site, and can rattle, while Cu2 (dark blue) is in square-planer coordinated $A'$ site. FeO$_6$ and ReO$_6$ octahedra are shown by yellow and green, respectively. In CuCu$_3$V$_4$O$_{12}$, both these octahedra are filled by V ions instead of Fe and Re.}
\end{figure}

Anharmonic vibrations of weakly bound ions in an oversized atomic cage formed by the other atoms are commonly known as  rattling. They have been observed in  materials such as VAl$_{10+\delta}$~\cite{Caplin1973}, clathrates~\cite{Eisenmann1986}, dodecaborides~\cite{Sluchanko2018}, filled skutterudites~\cite{Braun1980}, $\beta$-pyrochlore oxides~\cite{Yamaura2006}. Rattling or other types of anharmonicity can lead, e.g., to Schottky-type anomaly of specific heat at low temperature \cite{Hasegawa}, result in significant increase of electron effective mass \cite{Brihwiler2006,Grosche2001,Bauer2002}, suppress thermal conductivity~\cite{Jana2016,Chang2018} or be a driving force for the superconductivity~\cite{Brihwiler2006, Grosche2001, Bauer2002, Nagao2009}.

Recently, the rattling has been suggested for quadruple perovskite CuCu$_3$V$_4$O$_{12}$ synthesized under a high-pressure~\cite{Akizuki2015}. In quadruple perovskites AA'$_3$B$_4$O$_{12}$ the A site ions are icosahedrally (twelve neighbors) coordinated by oxygen atoms. The thermal displacement parameter of Cu ions at $A$ site in CuCu$_3$V$_4$O$_{12}$ was found to be quite large $U_{iso}$$\approx$0.045~\AA$^2$ at 300 K. Together with unusual behavior in specific heat $C_p(T)$ this led to suggestion of possible rattling in CuCu$_3$V$_4$O$_{12}$~\cite{Akizuki2015}.

In the present paper we report direct evidence of a rattling mode in  CuCu$_3$V$_4$O$_{12}$. The total energy density functional theory (DFT) calculations clearly show rattling distortions along [111] direction related to the Cu vibration. Moreover, there are local minima not only in this direction, but also in [001] and [110] ones, so that the effective potential is not of a simple double-well character, but has a complicated multi-well structure. This effect has been studied for another recently synthesized quadruple perovskite CuCu$_3$Fe$_2$Re$_2$O$_{12}$.

\section{Calculation details}

The electronic structure calculations were performed in the local density approximation taking into account Coulomb repulsion within LDA+U method in rotationally invariant form introduced by Liechtenstein {\textit {et al}} ~\cite{Liechtenstein95} using the Vienna {\textit {ab initio}} simulation package (VASP)~\cite{Kresse93}.

The experimental crystal structure data with the space groups No. 204 and No. 201 reported in Refs.~\cite{Akizuki2015} and~\cite{Long_str} for CuCu$_3$V$_4$O$_{12}$ and CuCu$_3$Fe$_2$Re$_2$O$_{12}$, respectively, were used. The cubic  Brillouin zone was sampled on a mesh of 6  x 6 x 6 {\boldmath $k$}~points. The cutoff energy for the plane-wave basis was set to 500 eV. The convergence criteria for the total energy calculations was set to 10$^{-7}$ eV. In the total energy calculations with displaced Cu ions other atoms were fixed at their initial positions. The values of the onsite Coulomb repulsion and Hund exchange parameters were taken close to what is used in the literature for corresponding transition metals: $U$(Cu) = 7.5~eV, $U$(V) = 3.5~eV, $U$(Fe) = 4.4~eV, $U$(Re) = 2.0~eV and $J_{\rm H}$(Cu) = 1.0~eV, $J_{\rm H}$(V)=0.7~eV, $J_{\rm H}$(Fe) = 0.9~eV, $J_{\rm H}$(Re) = 0.5~eV~\cite{Zakharov2014,Nekrasov2006,Pchelkina2013,Taran2023}.

Calculations for CuCu$_3$V$_4$O$_{12}$  were carried out with antiferromagnetic configuration of Cu2 ions, as well as antiferromagnetic arrangement of magnetic moments on nearest V atoms. The configuration with ferromagnetic order on Cu2 ions was found to be higher in total energy than the one with antiferromagnetic configuration. 
 For CuCu$_3$Fe$_2$Re$_2$O$_{12}$ the magnetic moments of Cu2 and Fe atoms were chosen to be ferromagnetically ordered. It is rather essential to expect strong superexchange interaction between Fe and Re atoms that gives antiferromagnetic order, that is why the Re magnetic moments were arranged in the opposite to the Cu2 and Fe magnetic moments direction.

\begin{figure}[h]
\includegraphics[width=0.85\columnwidth,angle=270]{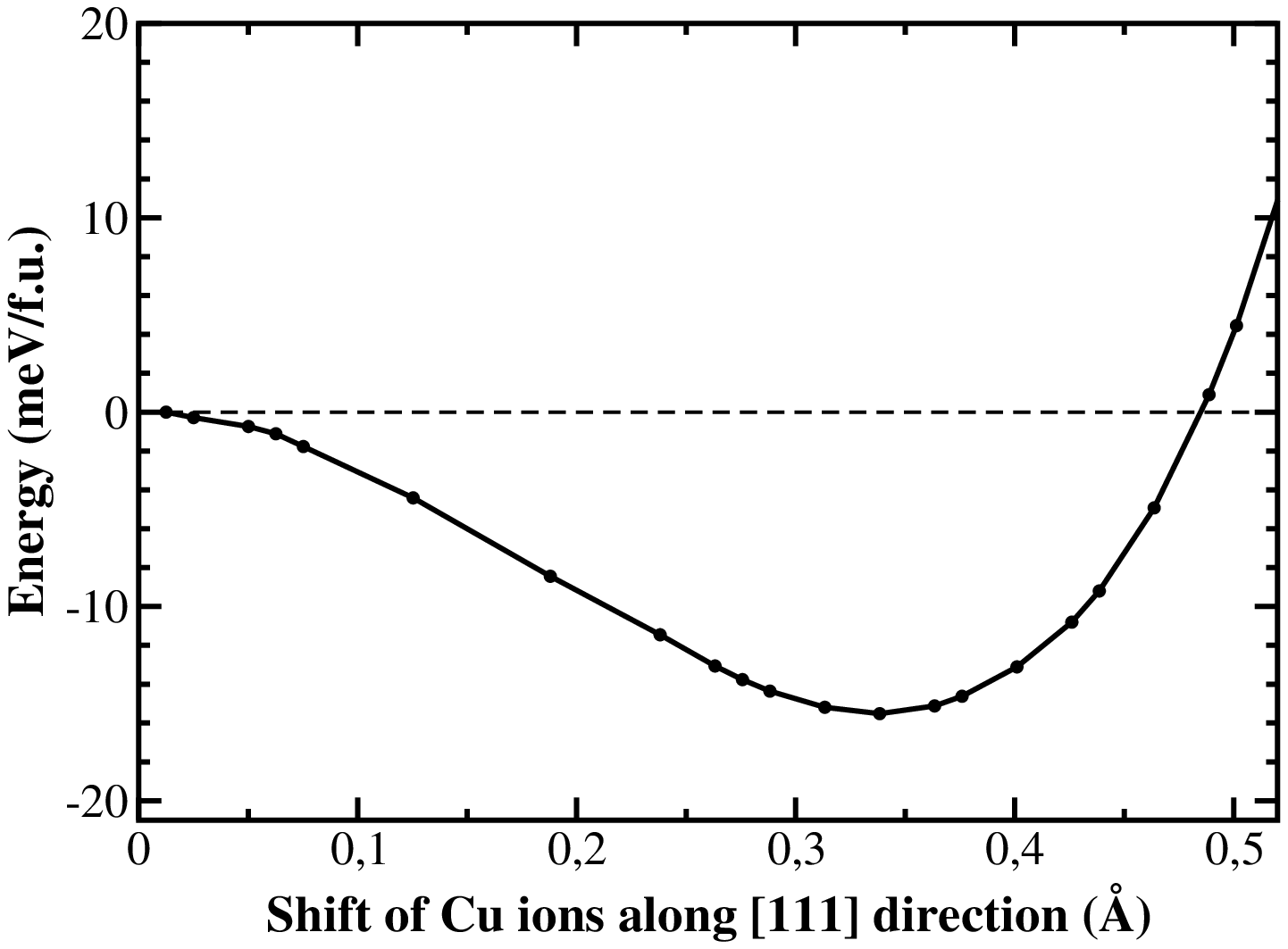}
\caption{\label{cucu3v} Total energy (per formula unit) vs Cu displacement in [111] direction in CuCu$_3$V$_4$O$_{12}$.}
\end{figure}

\section{Results: CuCu$_3$V$_4$O$_{12}$}  

The crystal structure of quadruple perovskites AA'$_3$B$_4$O$_{12}$ on example of CuCu$_3$Fe$_2$Re$_2$O$_{12}$ is shown in Fig.~\ref{cucufere_str}. CuCu$_3$V$_4$O$_{12}$ has a similar structure with V ions occupying both octahedral positions. The specific feature of quadruple perovskites is that $A$ sites in initial perovskite structure  must be (at least partially) occupied by Jahn-Teller active transition metals (TM). This leads to strong lattice distortions, formation of CuO$_4$ square planars, which helps to order $B$-site ions. Typically 3/4 of $A$ initial perovkite sites ($A'$ sites) are occupied by Jahn-Teller ions, while one can use rare-earths or other transition metal (TM) ions for remaining 1/4 positions ($A$ sites). 

CuCu$_3$V$_4$O$_{12}$ looks like an exemption since both $A$ sites are occupied by Cu ions. However, this is not the case - crystallographically there are two different Cu positions. Moreover, our calculations as well as previous results~\cite{Mehmood2022} show that Cu at $A$ sites (i.e. those, which are in icosahedra) have $3d^{10}$ electronic configuration, i.e. these are Cu$^{1+}$ ions. Cu ions at $A'$ site have local magnetic moment and these are Jahn-Teller active Cu$^{2+}$ ions. 

Distance between Cu$^{1+}$ and surrounding oxygens (2.548 \AA) is much larger than sum of corresponding ionic radii  $R_{VI}(Cu^{2+})+R_{VI}(O^{2-})$=0.73+1.4=2.13~\AA ~(ionic radius for Cu$^{1+}$ is expected to be even smaller, but it is absent in Shannon's data for large coordination numbers)~\cite{Shannon1976}. This is the reason why there develops a localized phonon mode with Cu vibrating in this large O$_{12}$ cage, a rattling mode. Fig.~\ref{cucu3v} shows how the total energy in DFT+U calculations depends on Cu$^{1+}$ displacement. One can notice a huge, $\sim$0.35\AA, distortion corresponding to the minimum of potential. This shift practically does not depend on choice of $U$ and $J_H$ parameters. Thus our calculations directly demonstrate presence of the rattling mode in CuCu$_3$V$_4$O$_{12}$. Moreover, this mode does not seem to be specific for this particular material and therefore we tested whether rattling vibrations present in another quadruple perovskite, namely, recently synthesized CuCu$_3$Fe$_2$Re$_2$O$_{12}$~\cite{Long_str}.



\section{Results: CuCu$_3$Fe$_2$Re$_2$O$_{12}$}

 Detailed analysis of the electronic and magnetic properties of this material will be published elsewhere, while here we would like to concentrate on the possible rattling distortions of Cu ion occupying $A$ site, labeled as Cu1 in Fig.~\ref{cucufere_str}. Moreover, we will study them in more details than in the case of CuCu$_3$V$_4$O$_{12}$.

\begin{figure}[h]
\includegraphics[width=0.95\columnwidth]{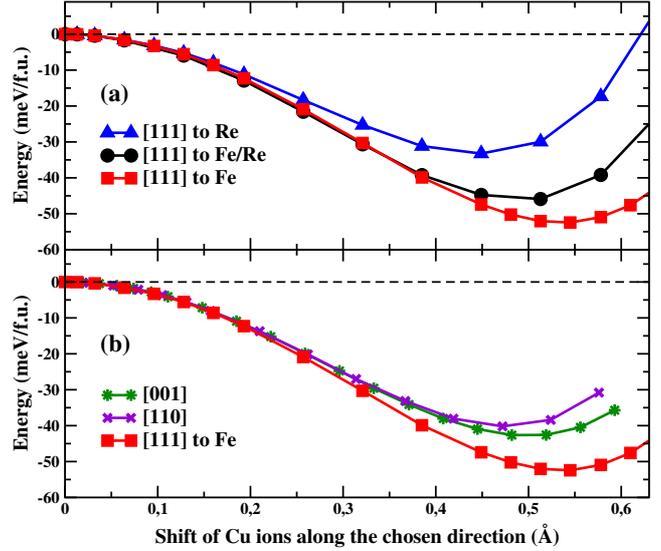}
\caption{\label{cucu3fere} Total energy $vs$ Cu1 displacement (a) along [111] direction, when two different Cu1 ions in a unit cell are shifted both to Re, one to Re and one to Fe, and both to Fe atoms, and (b) along [001], [110], and [111] directions in CuCu$_3$Fe$_2$Re$_2$O$_{12}$.}
\end{figure}

In fact, there are several possible types of rattling. Indeed, as one can see from Fig.~\ref{cucufere_str}, there are two Cu1 ions in the unit cell: those siting in the center of cube and in its corners. Therefore, rattling distortions of these two Cu1 ions can be in the same [111] direction, when the distance between them does not change (in-phase distortion; one of Cu1 goes to Fe and another to Re), or in the opposite directions when we have two different Cu1-Cu1 distances (out-of-phase distortion). Moreover, there are two inequivalent by symmetry out-of-phase distortions with Cu1 ions moving to Fe or Re ions between them (i.e., these are in fact [111] and [-1-1-1] distortions). Fig.~\ref{cucu3fere}a summarizes results of calculations for these 3 types of possible rattling distortions. One can see that this is the out-of-phase distortion to Fe ions, which gives the lowest total energy. The origin of this effect can be purely electrostatic: positive Cu1$^{1+}$ ion repulses stronger from Re$^{5.5+}$ than from Fe$^{3+}$.

Next, we checked other directions of possible rattling distortions and found that there are indeed local minima in [001], along [110] directions (and others equivalent by symmetry), but all of them have higher energies than the one along [111] direction, see Fig.~\ref{cucu3fere}b. On the one hand, this result seems trivial and can be naturally explained by pure geometrical arguments: there is simply ``less space'' for shifting Cu1 in [001] or [110] direction. On the other hand, it demonstrates that the potential for the rattling has a complex form with many local minima and can not be described by a simplified double-well shape.

As one can see, in general the obtained displacement amplitude for the Cu ion while rattling in CuCu$_3$Fe$_2$Re$_2$O$_{12}$ is about 0.45-0.55 \AA, that is by 0.1 \AA ~larger than for CuCu$_3$V$_4$O$_{12}$. It can be explained by difference in the unit cell volumes and, consequently, Cu-TM bond lengths for the initial undistorted structure. Cu$^{1+}$ is in the center of the cube formed by TM ions. In the case of CuCu$_3$Fe$_2$Re$_2$O$_{12}$, the edge of the cube is 3.705 \AA~while it is 3.618 \AA~for the CuCu$_3$V$_4$O$_{12}$. Moreover, the distance between Cu1 and surrounding oxygens is even larger than for the previously discussed compound (2.763 \AA). In addition, the energy gain for CuCu$_3$Fe$_2$Re$_2$O$_{12}$ is also 2-3 times larger (30-50 meV/f.u. $vs.$ 15 meV/f.u.).

While amplitude of rattling is rather large, it does not affect magnetic moments and valence states of Cu, Fe, and Re. For the initial structure with the Cu ions exactly in the centre of cage the magnetic moments were found to be $\mu_{0}^{Cu2}$=0.48~$\mu_B$, $\mu_{0}^{Fe}$=4.06~$\mu_B$, and $\mu_{0}^{Re}$=$-0.70$~$\mu_B$. In the structure corresponding to the energy minimum along [111] direction they are practically the same being 
$\mu_{min}^{Cu2}$=0.49~$\mu_B$, $\mu_{min}^{Fe}$=4.05~$\mu_B$, and $\mu_{min}^{Re}$=-0.73~$\mu_B$.

\section{Conclusions} 

Analyzing our DFT+U calculation results concerning dependence of the total energy on the Cu1 ($A$-site) displacement from the center of O$_{12}$ icosahedral cage, we provide theoretical evidence of the presence of rattling modes in quadruple perovskite CuCu$_3$V$_4$O$_{12}$. We predict the same phenomenon in the recently synthesised CuCu$_3$Fe$_2$Re$_2$O$_{12}$. The [111] direction of Cu1 rattling turns out to be the most energetically favourable. In addition, dependence of the rattling amplitude and energy gain on the volume was proved. Therefore one can argue that pressure can suppress rattling modes. 

There are other atoms that may occupy the A site position in the quadruple perovskites, e.g. Yb, Ga, or La. It would be interesting and useful to check both theoretically and in the experiment whether rattling takes place there or not. Taking into account just corresponding ionic radii one would expect rattling is completely suppressed there or has a much smaller amplitude. 

Strong anharmonicity of the lattice potential, including its double-well or even multi-well behavior, and presence of the localized (rattling) phonon modes can result in a number of interesting effects in electronic properties and possibly provide  superconductivity mechanisms. In particular, an analysis within a simple harmonic approximation with an Einstein mode demonstrated anomalous specific heat in CuCu$_3$V$_4$O$_{12}$: the Sommerfeld coefficient was found to be extremely large, $\gamma$  = 126 mJ mol$^{-1}$ K$^{-2}$~\cite{Akizuki2015}. Note that a similar enhancement of electronic effective mass occurs in $\beta$-pyrochlore KOsO$_6$~\cite{Brihwiler2006}, and filled-skutterudite compounds PrOs$_4$Sb$_{12}$~\cite{Bauer2002}, NdOs$_4$Sb$_{12}$~\cite{Ho} and SmOs$_4$Sb$_{12}$~\cite{Sanada}. We argue that a similar enhancement of specific heat due to rattling can be observed in other quadruple perovskites. Moreover, one might expect anomalous suppression of the thermal conductivity  due to rattling, since low-energy vibrations serve as efficient scattering centers (cf. \cite{Bauer,Baumbach}).

{\bf Acknowledgements.} The work was supported by the Russian Science Foundation via project RSF 23-42-00069, and the National Natural Science Foundation of China (Grant No 12261131499).

\end{document}